\documentclass{basi}
\usepackage{epsfig}
\usepackage{amsmath}
\usepackage{amssymb}
\def\Q{\ifmmode\mathcal{Q}\else$\mathcal{Q}$\fi}
\newcommand{\msun}{M$_{\odot}$}
\newcommand\ion[2]{#1$\;${\small\rmfamily\@{#2}}\relax}%

\begin{document}
\title[The structures of embedded clusters]{The structures of embedded clusters} 
\author[S. Schmeja]%
       {Stefan Schmeja\thanks{e-mail: sschmeja@ita.uni-heidelberg.de} \\ 
        Zentrum f\"ur Astronomie der Universit\"at Heidelberg, 
              Institut f\"ur Theoretische Astrophysik,\\Albert-Ueberle-Str.~2, 
                  69120 Heidelberg, Germany}

\maketitle

\label{firstpage}

\begin{abstract}
Stars are usually formed in clusters in the dense cores of molecular clouds.
These embedded clusters show a wide variety of morphologies from hierarchical clusters with substructure
to centrally condensed ones. 
Often they are elongated and surrounded by a low-density stellar halo.
The structure of an embedded cluster, i.e.\ the spatial distribution of its members, 
seems to be linked to the complex structure of the parental molecular cloud and 
holds important clues about the formation mechanism and the initial conditions,
as well as about the subsequent evolution of the cluster.
\end{abstract}


\section{Introduction}
\label{sec:intro}

Stars are born in the dense cores of turbulent molecular clouds that fragment
and collapse under their own gravity.
Once a gas clump becomes gravitationally unstable,
it begins to collapse and the central density
increases considerably until a proto\-stellar
core forms in the centre, which grows in mass
by accretion from the infalling envelope.
It is believed that the star formation process is controlled by the interplay between gravity
and supersonic turbulence (Mac Low \& Klessen 2004; Ballesteros-Paredes et al.\ 2007; McKee \& Ostriker
2007).
Supersonic turbulence, which is observed ubiquitously in molecular clouds with typical Mach numbers in the range
$1 \lesssim \mathcal{M} \lesssim 10$, plays a dual r\^ole.
While on large scales it supports clouds against contraction, on small scales it can
lead to local overdensity and collapse.

A large fraction of stars in the Milky Way form in clusters and aggregates of various size and mass 
(Pudritz 2002; Lada \& Lada 2003; Allen et al.\ 2007), appearing to form a hierarchy of structures.
The interstellar medium shows a hierarchical structure (sometimes described as fractal) from the largest 
giant molecular cloud (GMC) scales down to individual cores and clusters, which are sometimes hierarchical
themselves. There is no obvious change in morphology at the cluster boundaries, so clusters 
can be seen as the bottom parts of the hierarchy, where stars have had the chance to mix
(Efremov \& Elmegreen 1998; Elmegreen et al.\ 2000; Elmegreen 2009).
The complex hierarchical structure of turbulent molecular clouds provides a natural explanation 
for clustered star formation.
Molecular clouds vary enormously in size and mass. In small, low-density clouds stars form with low efficiency,
more or less in isolation or in distributed small groups of up to a few dozen members. 
Denser and more massive clouds may build up associations and clusters of several thousand stars.
The latter is rare in the solar neighbourhood, but observed in more extreme environments like the
Galactic centre or interacting galaxies.

Young clusters usually contain young stellar objects (YSOs) belonging to different evolutionary
classes, from deeply embedded Class~0 and Class~1 protostars to relatively
evolved pre-main-sequence stars (Class~2 and 3).
The morphologies of embedded clusters show a wide variety.
There are two basic types of clusters with regard to their structure (Lada \& Lada 2003):
(1) hierarchical clusters showing a stellar surface density distribution with
multiple peaks and possible fractal substructure; (2) centrally condensed
clusters exhibiting highly centrally concentrated stellar surface density distributions with
relatively smooth radial profiles that can be approximated by simple power-law functions 
or King (1962) profiles.

The analysis of cluster structures requires a complete and unbiased sample of the cluster members.
Since embedded stars are best visible in the infrared part of
the spectrum, and since nearby GMCs can span several degrees
on the sky, wide-field near and mid-infrared surveys are useful tools 
for identifying and mapping the distribution of young stars in GMCs.
Consequently, recent infrared surveys such as 2MASS or several {\em Spitzer Space Telescope}
surveys have revealed a great deal about embedded clusters and their structures.
On the other hand, numerical simulations are able to predict the formation of a stellar cluster
and its properties with increasing reliability.
This review attempts to summarise recent observational and theoretical findings about cluster
structures and in particular their clues to star formation and early cluster evolution.

\section{Methods to analyse clustering}
\label{sec:methods}

A straightforward way to investigate cluster structures is to study the stellar density distribution
derived from simple star counts (e.g.\ Lada \& Lada 1995) or the nearest neighbour (NN) density
(Casertano \& Hut 1985).
These methods allow to identify clusters as density enhancements in a field as well as the study of the
internal density distribution.
By varying the value of the $j$th nearest neighbour, the NN method can account for
the large-scale structure (using large $j$) and small-scale density fluctuations (small $j$).

The mean surface density of companions (MSDC; Larson 1995) $\Sigma(\theta)$ specifies the average number of neighbours per square degree on the sky at an angular separation $\theta$ for each cluster star.
Knowing the distance to the cluster, $\theta$ can be converted to an absolute distance $r$ to
determine $\Sigma(r)$.

The parameter \Q\ (Cartwright \& Whitworth 2004, 2009a) allows to quantify the clustering and to 
distinguish between a radial density distribution and fractal subclustering.
It is defined as $\Q = \ell_{\rm MST}/\bar{s}$, combining the normalized correlation length
$\bar{s}$ (the mean distance between all cluster stars), and the normalized mean edge length of
the minimum spanning tree $\ell_{\rm MST}$ of the stars.
Large \Q\ values ($\Q > 0.8$) describe centrally condensed clusters
having a volume density $n(r) \propto r^{-\alpha}$, while small values
($\Q < 0.8$) indicate clusters with fractal substructure.
\Q\ is correlated with the radial density exponent $\alpha$ for
$\Q > 0.8$ and anticorrelated with the fractal dimension $D$ for $\Q < 0.8$.
The method is now widely used and has been applied successfully to observed embedded and open
clusters as well as to the results of numerical simulations 
(e.g.\ Schmeja \& Klessen 2006; Schmeja et al.\ 2008b; Gieles et al.\ 2008; Bastian et al.\ 2009; S\'anchez \& Alfaro 2009).

\section{Cluster morphology}

\begin{figure}[t]
\includegraphics[width=\textwidth]{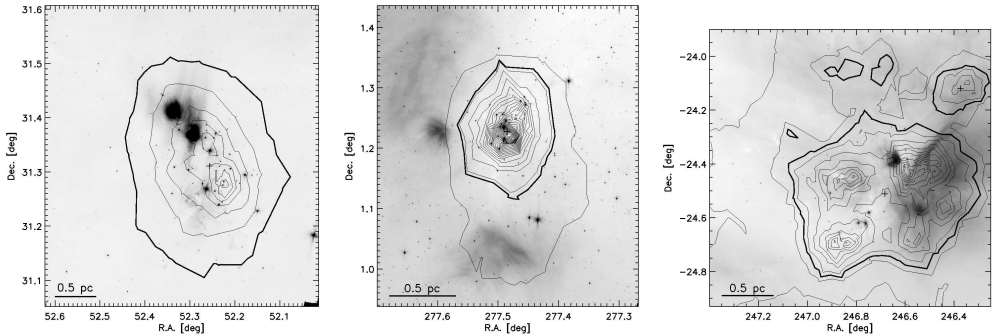}
   \caption{{\em Spitzer} IRAC 8\,$\mu$m images of the three embedded clusters NGC~1333 (left; $\Q=0.81$),
    Serpens~A (middle; $\Q=0.84$) and L1688 in Ophiuchus (right; $\Q=0.72$)
   overlaid with the 20th NN density contours. The thick contours correspond to the cluster boundaries 
   defined as $3 \sigma$ above the background level. (See text for details. Figure adapted from Schmeja et al.\ 2008b.)}
   \label{fig:c2dclusters}
\end{figure}

Only few embedded clusters appear spherically symmetric, most of them have an elongated 
shape (e.g.\ Allen et al.\ 2007; see also Fig.~\ref{fig:c2dclusters}).
Many young clusters have a clumpy structure and show two or more pronounced 
density peaks or subclusters (and values $\Q < 0.8$). 
The shape of young clusters and the distribution of its YSOs are often aligned with the filamentary structure
seen in the parental molecular cloud (Hartmann 2002; Palla \& Stahler 2002; Gutermuth et al.\ 2005; 
Teixeira et al.\ 2006; Schmeja et al.\ 2008b).
The elongated and clumpy morphologies probably reflect the initial structure of the dense 
molecular gas in the cloud.
Circularly symmetric clusters, on the other hand, may be explained as a 
result of the expansion of the cluster following the dissipation of the molecular gas
(Gutermuth et al.\ 2005).

Young clusters usually show a low-density halo surrounding the dense cluster cores
(e.g.\ Ma{\'{\i}}z-Apell{\'a}niz 2001; Muench et al.\ 2003; Ascenso et al.\ 2007), 
and a smooth transition to the distributed YSO population
spread throughout the entire molecular cloud, making the exact definition of cluster
boundaries difficult if not impossible.
Whether the observed halos have been formed in situ or are a result of dynamical evolution
is not obvious, as there is both evidence for continuing star formation in halos 
(Megeath et al.\ 2004) and a halo population that is older than the one in the core 
(Ascenso et al.\ 2007).

\section{The spatial distribution of young stars}

The spacing of protostars is an important constraint on
the mechanisms of cloud fragmentation and possible
subsequent interactions by protostars.
Typical separations of protostars in embedded clusters
 are found in the range between about 0.02 and 0.10~pc
(Teixeira et al.\ 2006; Schmeja et al.\ 2008b; Gutermuth et al.\ 2008, 2009).
Gutermuth et al.\ (2009) found a relatively similar spacing of protostars of $0.072 \pm 0.006$~pc
in 39 nearby embedded clusters. The value of 0.072~pc corresponds to the Jeans
length in an isothermal medium with a temperature of 20~K and a density of 
$2 \times 10^5 \rm{cm}^{-3}$, properties typical for the dense cores of 
star-forming molecular clouds. 
These values are therefore interpreted as a a signature
of Jeans fragmentation, in agreement with the conclusions by Larson (1995) and Teixeira et al.\ (2006).

The MSDC in nearby embedded clusters is best described by a double power law, 
with slopes of $\beta \approx -2$ for small
separations and  $-0.9 < \beta < -0.1$ for larger separations (Larson 1995; 
Nakajima et al.\ 1998; Bate et al.\ 1998; Gladwin et al.\ 1999; Klessen \& Kroupa
2001), suggesting two distinct clustering regimes. 
At small separations, the MSDC is determined by
binaries and higher-order multiple systems, whereas at large separations it reflects the overall 
spatial structure of the cluster.
In Taurus-Auriga the break of the distribution is found at $\sim 0.04$\,pc,  
which is equivalent to the typical Jeans length in molecular clouds, corresponding to a Jeans mass
of $\sim 1$~\msun\ (Larson 1995), suggesting that stellar systems with smaller
separations form from the fragmentation of single collapsing
protostellar cores, whereas the spatial distribution on larger scales follows the hierarchical structure of the
parental molecular cloud.
However, the length scale of the break of
the distribution varies significantly for different clusters, between about 0.002 and 0.15~pc.
This casts doubt on the interpretation of the break length as determined by the Jeans criterion,
as this would imply fairly different Jeans masses and in turn strong deviations in the initial 
mass function, which are not observed.

In many clusters the spacing and distribution of protostars (Class~0/1) is significantly
different from that of Class~2/3 pre-main-sequence stars (Palla \& Stahler 2002; 
Kaas et al.\ 2004; Winston et al.\ 2007; Muench et al.\ 2007).
The young Class~0/1 objects, which had less time to move away from their birth sites,
 tend to be concentrated in compact small groups, often associated with 
clumps and filaments of molecular gas and dust,
while the older Class~2/3 sources are spread out
over a more extended region (see also \S\ref{evolution}).

Brown dwarfs seem to follow a spatial distribution different from stars, at least in some clusters.
While the stellar population in IC~348 and the Trapezium cluster displays a centrally concentrated
distribution, the brown dwarfs are distributed more homogeneously in space within twice the
cluster radius, in agreement with the scenario of brown dwarfs formed as ``ejected embryos''
 from multiple systems (Kumar \& Schmeja 2007).

\section{Dynamical evolution}
\label{evolution}

Numerical simulations of gravoturbulent cloud fragmentation predict that clusters 
build up in a hierarchical way from subclusters which evolve dynamically and merge together
to form the final, centrally condensed cluster (e.g.\ Bonnell et al.\ 2003; Bonnell \& Bate 2006).
Schmeja \& Klessen (2006) investigated the temporal evolution of a large set of numerical 
simulations of gravoturbulent fragmentation and showed that the \Q\ parameter of all resulting
clusters increases with time.
This is confirmed by observations of embedded clusters in the Perseus, Serpens and Ophiuchus molecular clouds, 
where the younger Class~0/1 sources show substantially lower \Q\ values than the more evolved Class~2/3 objects 
(Schmeja \& Klessen 2006; Schmeja et al.\ 2008b).
Substructure in a fractal cluster may be erased rapidly or preserved for a longer time, 
depending on the velocity dispersion.
According to N-body simulations (Goodwin \& Whitworth 2004), in clusters with a low initial velocity 
dispersion, the resulting collapse of the cluster tends to erase substructure to a large extent.
In clusters with virial ratios of 0.5 or higher, however, initial substructure survives for
several crossing times.
Spatial substructure has been observed in clusters as old as $\sim$100~Myr (S\'anchez \& Alfaro 2009).
However, the structure of open clusters may also be a result of later dynamical evolution.
The investigation of a large sample of embedded and open clusters 
shows that, although open and embedded clusters seem to cover roughly the same range in \Q,
the mean \Q\ value of the sample of embedded clusters is larger than
the mean \Q\ value of the open clusters (Schmeja et al.\ 2008a).
This may suggest that a cluster evolving from the embedded phase to an open cluster
regresses to a more hierarchical configuration due to dynamical interactions, tidal fields etc.
The central condensation may be most significant in the late embedded phase, when gravity is the
dominant force.

As the stellar velocity dispersion may be linked to the velocity dispersion of the gas, 
the structure of an embedded cluster may be related to the turbulent energy in a molecular cloud in a 
way that clusters in regions with low Mach numbers reach a centrally condensed configuration much
faster than those in highly turbulent clouds (Schmeja et al.\ 2008b).

Many embedded (and open) clusters show signs of mass segregation, i.e.\
the brightest, most massive cluster members are concentrated
toward the centre of the cluster.\footnote{Note however concerns 
that the commonly used indicators for mass segregation
are severely biased by statistical and incompleteness effects, so that
there may be no evidence for mass segregation in young clusters at all
(Ascenso et al.\ 2009).}
Whether mass segregation is primordial (stars in the cluster centre accrete more material
due to their location at the bottom of the cluster potential well) or
a result of dynamical evolution (massive stars formed elsewhere 
in the cluster eventually sink to the centre through the effects of two-body relaxation)
is not clear.
The fact that mass segregation is observed even in very young clusters (e.g.\ 
Gouliermis et al.\ 2004; Chen et al.\ 2007; Sharma et al.\ 2007)
may be seen in favour of the primordial scenario, but simulations show that
dynamical mass segregation can occur on very short timescales (Allison et al.\ 2009).
Schmeja et al.\ (2008b) investigated the variation of \Q\ as a function of magnitude
(and thus, mass) for several embedded clusters and found a 
decrease of \Q\ with fainter magnitudes in the older (2-3~Myr) clusters
and a constant or reverse relation in the youngest clusters,
indicating that the effect of mass segregation through dynamical interaction becomes visible at
a cluster age between about 2 and 3~Myr.

\section{Effects of massive stars and triggered star formation}

Massive OB stars are part of many large star-forming regions, emitting
strong UV radiation and driving stellar winds. This feedback process can seriously
affect the surrounding molecular gas, preventing the formation or growth of stars in their
vicinity, while simultaneously triggering star formation at the periphery of the cloud
(e.g.\ Elmegreen \& Lada 1977; Whitworth et al.\ 1994; Zinnecker \& Yorke 2007).
This obviously has a strong influence on the cluster structure and may lead to an anti-clustered distribution
of YSOs, as it is observed e.g.\ in the OB-stars driven dust bubble CN138
(Watson et al.\ 2009; Cartwright \& Whitworth 2009b), the W5 star-forming region (Koenig et al.\ 2008),
or around the ionized bubble RCW~120 (Deharveng et al.\ 2009).

Embedded clusters are sometimes found aligned with expanding \ion{H}{II} shells, e.g.\ in the Rosette
Molecular Cloud (Phelps \& Lada 1997), suggesting that their formation was triggered by external compression.
The large star-forming region NGC~346 in the Small Magellanic Cloud shows a filamentary structure
of increased stellar density with several clusters of pre-main-sequence stars that coincide
with an arc-like structure of gas and dust (Hennekemper et al.\ 2008; Schmeja et al.\ 2009).
This distribution is interpreted as a signature of triggered star formation by a 
wind-driven expanding \ion{H}{II} region (or bubble) blown
by a massive supernova progenitor and possibly other bright stars 
a few Myr earlier (Gouliermis et al.\ 2008).

\section{Concluding remarks}

Embedded clusters show a wide variety of structures, common features are an elongated
shape, substructure, and a low-density halo. Clusters seem to evolve from an initial hierarchical
configuration consisting of several subclusters toward a centrally condensed state.
Therefore the structures of embedded clusters hold important clues to their formation mechanism, and, in
a later stage, reflect their dynamical evolution.
Furthermore, they are an observational constraint to the results of numerical simulations 
and the knowledge of a cluster's structure can help identifying new cluster members
and discriminating cluster members from field stars (albeit only in a statistical sense).
While recent infrared surveys and hydrodynamical and N-body simulations have
significantly improved our understanding of the structures of young clusters, there are
still a lot of open questions, in particular when it comes to massive clusters or clusters
in galaxies other than our Milky Way.

\section*{Acknowledgements}
I am grateful to Joana Ascenso, Nate Bastian, Annabel Cartwright, Dirk Froebrich, Dimitrios Gouliermis, 
Ralf Klessen, Nanda Kumar, and Paula Stella Teixeira for helpful discussions on cluster structures.
I also wish to thank Devendra Ojha and the organisers for an inspiring workshop and for their hospitality.
My work is supported by the {\em Deutsche Forschungs\-ge\-meinschaft} (DFG) through grant
SCHM~2490/1-1.

\label{lastpage}

\begin{thebibliography}{10}

\bibitem{allen07}
Allen L., et al., 2007,
  in Protostars and Planets~V, eds.\ B. Reipurth, D. Jewitt, \& K. Keil, Tucson: Univ.\ of Arizona Press, 361.

\bibitem{allison09}
Allison R.~J., et al., 2009, ApJ, 700, L99.

\bibitem{ascenso07}
Ascenso J., et al., 2007, A\&A, 476, 199.

\bibitem{ascenso09}
Ascenso J., Alves J., \& Lago M.~T.~V.~T., 2009, A\&A, 495, 147.

\bibitem{bp07}
Ballesteros-Paredes J., et al., 2007,
  in Protostars and Planets~V, eds.\ B. Reipurth, D. Jewitt, \& K. Keil, Tucson: Univ.\ of Arizona Press, 63.

\bibitem{bastian09B}
Bastian N., et al., 2009, MNRAS, 392, 868.

\bibitem{bate98}
Bate M.~R., Clarke C.~J., \& McCaughrean M.~J., 1998, MNRAS, 297, 1163.

\bibitem{bonnell+bate06}
Bonnell I.~A., \& Bate M.~R., 2006, MNRAS, 370, 488.

\bibitem{bonnell03}
Bonnell I.~A., Bate M.~R., \& Vine S.~G., 2003, MNRAS, 343, 413.

\bibitem{cw04}
Cartwright A., \& Whitworth A.~P., 2004, MNRAS, 348, 589.

\bibitem{cw09a}
Cartwright A., \& Whitworth A.~P., 2009a, MNRAS, 392, 341.

\bibitem{cw09b}
Cartwright A., \& Whitworth A.~P., 2009b, A\&A, 503, 909.

\bibitem{casertano+hut85}
Casertano S., \& Hut P., 1985, ApJ, 298, 80.

\bibitem{chen07}
Chen L., de Grijs R., \& Zhao J.~L., 2007, AJ, 134, 1368.

\bibitem{deharveng09}
Deharveng L., et al., 2009, A\&A, 496, 177.

\bibitem{efremov+elmegreen98}
Efremov Y.~N., \& Elmegreen B.~G., 1998, MNRAS, 299, 588.

\bibitem{elmegreen06}
Elmegreen B.~G., 2009, 
   in Globular Clusters -- Guides to Galaxies, eds.\ T. Richtler \& S. Larsen, 
   ESO Astrophysics Symposia, Berlin/Heidelberg: Springer, 87.

\bibitem{elmegreen+lada77}
Elmegreen B.~G., \& Lada C.~J., 1977, ApJ, 214, 725.

\bibitem{elmegreen00}
Elmegreen B.~G., et al., 2000, 
   in Protostars and Planets~IV, eds.\ V. Mannings, A.~P. Boss, \& S.~S. Russell, 
   Tucson: Univ.\ of Arizona Press, 179.

\bibitem{gieles08}
Gieles M., Bastian N., \& Ercolano B., 2008, MNRAS, 391, L93.

\bibitem{gladwin99}
Gladwin P.~P., et al., 1999, MNRAS, 302, 305.

\bibitem{gw04}
Goodwin S.~P., \& Whitworth A.~P., 2004, A\&A, 413, 929.

\bibitem{gouliermis04}
Gouliermis D.~A., et al., 2004, A\&A, 416, 137.

\bibitem{gouliermis08}
Gouliermis D.~A., et al., 2008, ApJ, 688, 1050.

\bibitem{gutermuth05}
Gutermuth R.~A., et al., 2005, ApJ, 632, 397.

\bibitem{gutermuth08}
Gutermuth R.~A., et al., 2008, ApJ, 674, 336.

\bibitem{gutermuth09}
Gutermuth R.~A., et al., 2009, ApJS, 184, 18.

\bibitem{hartmann02}
Hartmann L., 2002, ApJ, 578, 914.

\bibitem{hennekemper08}
Hennekemper E., et al., 2008, ApJ, 672, 914. 

\bibitem{kaas04}
Kaas A.~A., et al., 2004, A\&A, 421, 623.

\bibitem{king62}
King I., 1962, AJ, 67, 471.

\bibitem{klessen+kroupa01}
Klessen R.~S., \& Kroupa P., 2001, A\&A, 372, 105.

\bibitem{koenig08}
Koenig X.~P., et al., 2008, ApJ, 688, 1142.

\bibitem{kumar+schmeja07}
Kumar M.~S.~N., \& Schmeja S., 2007, A\&A, 471, L33.

\bibitem{lada+lada95}
Lada E.~A., \& Lada C.~J., 1995, AJ, 109, 1682.

\bibitem{lada+lada03}
Lada C.~J., \& Lada E.~A., 2003, ARA\&A, 41, 57.

\bibitem{larson95}
Larson R.~B., 1995, MNRAS, 272, 213.

\bibitem{maclow+klessen04}
Mac Low M.-M., \& Klessen R.~S., 2004, Rev.\ Mod.\ Phys., 76, 125.

\bibitem{maiz01} 
Ma{\'{\i}}z-Apell{\'a}niz J., 2001, ApJ, 563, 151.

\bibitem{mckee+ostriker07}
McKee C.~F., \& Ostriker E.~C., 2007, ARA\&A, 45, 565.

\bibitem{megeath04}
Megeath S.~T., et al., 2004, ApJS, 154, 367.

\bibitem{muench03}
Muench A.~A., et al., 2003, AJ, 125, 2029.

\bibitem{muench07}
Muench A.~A., et al., 2007, AJ, 134, 411.

\bibitem{nakajima98}
Nakajima Y., et al., 1998, ApJ, 497, 721.

\bibitem{palla+stahler02}
Palla F., \& Stahler S.~W., 2002, ApJ, 581, 1194.

\bibitem{phelps+lada97}
Phelps R.~L., \& Lada E.~A., 1997, ApJ, 477, 176.

\bibitem{pudritz02}
Pudritz R.~E., 2002, Science, 295, 68.

\bibitem{sancehz+alfaro09}
S\'anchez N., \& Alfaro E.~J., 2009, ApJ, 696, 2086.

\bibitem{sk06}
Schmeja S., \& Klessen R.~S., 2006, A\&A, 449, 151.

\bibitem{skfk08}
Schmeja S., et al., 2008a, 
in Dynamical Evolution of Dense Stellar Systems, eds.\ E. Vesperini, M. Giersz, \& A. Sills,
Proc.\ IAU Symp.\ 246, 50.

\bibitem{skf08}
Schmeja S., Kumar M.~S.~N., \& Ferreira B., 2008b, MNRAS, 389, 1209.

\bibitem{skf09}
Schmeja S., Gouliermis D.~A., \& Klessen R.~S., 2009, ApJ, 694, 367.

\bibitem{sharma09}
Sharma S., et al., 2007, MNRAS, 380, 1141.

\bibitem{teixeira06}
Teixeira P.~S., et al., 2006, ApJ, 636, L45.

\bibitem{watson09}
Watson C., et al., 2009, ApJ, 694, 546.

\bibitem{whitworth94}
Whitworth A.~P., et al., 1994, MNRAS, 268, 291.

\bibitem{winston07}
Winston E., et al., 2007, ApJ, 669, 493.

\bibitem{zinnecker+yorke07}
Zinnecker H., \& Yorke H.~W., 2007, ARA\&A, 45, 481.

\end{thebibliography}
\end{document}